# Isotopic disorder in Ge single crystals probed with $^{73}$Ge NMR

S. V. Verkhovskii

Institute of Metal Physics RAS, Ekaterinburg 620066, Russian Federation

A. Yu. Yakubovsky

Ecole Supérieure de Physique et de Chimie Industrielles, Paris, France and

Russian Research Centre "Kurchatov Institute", Moscow 123182, Russian Federation

B. Z. Malkin

Kazan State University, Kazan 420008, Russian Federation

S. K. Saikin

Clarkson University, Potsdam, NY 13699, USA and

Kazan State University, Kazan 420008, Russian Federation

M. Cardona

Max-Planck Institut für Festkörporperforschung, D-70569 Stuttgart, Germany

A. Trokiner

Ecole Supérieure de Physique et de Chimie Industrielles, Paris, France

V. I. Ozhogin

Russian Research Centre "Kurchatov Institute", Moscow 123182, Russian Federation

NMR spectra of $^{73}$Ge (nuclear spin $I = 9/2$) in germanium single crystals with different isotopic compositions have been measured at the frequency of 17.4 MHz at room temperature. Due to the small concentration (~0.1%) of the magnetic ($^{73}$Ge) isotope, the magnetic dipole-dipole interaction is negligible in the samples studied, and the observed specific features of the resonance lineshapes (a narrow central peak and a wide plateau) are determined mainly by the quadrupole interaction of magnetic nuclei with the random electric field gradient (EFG) induced by the isotopic disorder. The second and fourth moments of the distribution function of the EFG are calculated taking into account local lattice deformations due to mass defects in the close neighborhood of the magnetic nuclei, as well as charge density redistributions and lattice strains induced by distant impurity isotopes. The simulated lineshapes, represented by a superposition of Gaussians corresponding to individual transitions between nuclear Zeeman sublevels, agree reasonably well with the measured spectra.

## I. INTRODUCTION

The elemental semiconductors (diamond, silicon, germanium) contain different stable isotopes, and thus constitute the simplest type of a disordered solid. All physical properties of crystals depend, to some degree, on their isotopic composition. Detailed reviews on isotopic effects in the lattice dynamics of elemental semiconductors with the diamond structure and of binary compounds with the zincblende structure have been published recently[1,2,3]. Raman light scattering has been the main experimental method of studying frequencies and lifetimes of phonons in isotopically disordered crystals, although some inelastic scattering work has been reported[4]. Comprehensive studies of Raman scattering in germanium and silicon crystals with different isotopic composition are presented in Refs.[5,6]. The thermal conductivity and the heat capacity of the isotopically enriched Ge samples have been reported in Refs.[7,8].

Isotopic disorder affects the primary interatomic interactions through the changes of inter-ion distances only. The influence of the isotopic composition on the lattice constant can be described as the combined effect of the third order anharmonic terms in the potential energy of the lattice expanded in terms of displacements of atoms from their equilibrium positions and of the correlation functions of dynamic atomic displacements which depend on atomic masses[9]. A thermodynamic description of this effect based on Gruneisen parameters has also been used[10]. The resulting change in the lattice constant, like the shift of the phonon frequencies, depends on the average isotope mass of the corresponding atoms. Another type of isotopic effect is produced by the random distribution of isotopes over the crystal lattice. This randomness causes local fluctuations of a crystal density and local lattice deformations, which perturb the translational invariance of a crystal lattice. In the disordered lattice, some selection rules, based on phonon quasimomentum conservation and the symmetry properties of the perfect lattice, are lifted. In particular, first order Raman scattering by phonons throughout the Brillouin zone may be observed, and nonvanishing components of the electric field gradient (EFG) appear at the sites of the lattice with nominally cubic symmetry.

The random lattice strains may be studied by measuring NMR in the magnetic isotopes with the nuclear spin $I > 1/2$ and, correspondingly, with nonzero nuclear quadrupole moment. Impurity centers disturb the distribution of the electronic and nuclear charges and induce long-range lattice strains with an $r^{-3}$ dependence on the distance from the center. Both, the charge redistribution and the lattice distortions bring about random variations of the EFG at nuclear quadrupole moments. The structural information in disordered solids that may be extracted from the distribution of the EFG was already considered more than 40 years ago. The inhomogeneous quadrupolar broadening and profile changes of NMR lines were used, in particular, for quantitative studies of the lattice strains due to solute ions in alkali halide crystals with substitutional isovalent impurities[11][12] and for estimations of concentration of vacancy type defects in GaAs[13].

Remarkable changes of the $^{73}$Ge NMR lineshape, depending on the isotopic composition of a germanium crystal and on orientation of the external magnetic field relative to the crystallographic axes, were observed in previous work[14]. They suggest the possibility of monitoring the relative role of two main mechanisms of line broadening (magnetic dipole-dipole and electric quadrupole interactions) in crystals of high purity by varying the isotopic composition of the samples. In the perfect monoisotopic germanium crystal, the EFG is zero at the lattice sites with $T_d$ symmetry. The lattice strains due to mass fluctuations in the isotopically disordered sample induce random non-zero field gradients and cause the quadrupole broadening of magnetic dipole transitions between Zeeman sublevels of $^{73}$Ge nuclei (with nuclear spin $I = 9/2$).

In the present work, we compare experimental data taken on samples with different degrees of isotopic disorder and with the ineffective magnetic dipole-dipole broadening mechanism due to a very small concentration of magnetic $^{73}$Ge nuclei. A theoretical discussion of the NMR lineshape and of the quadrupole broadening induced by isotopic deformations in the crystal lattice is given. Both random EFG at the magnetic nuclei and random lattice strains are included in the framework of the anharmonic model of the adiabatic bond charges[15][16].

## II. THEORY OF QUADRUPOLE BROADENING OF THE $^{73}$GE NMR LINE IN THE ISOTOPICALLY DISORDERED AND MAGNETICALLY DILUTED GERMANIUM CRYSTALS

The Hamiltonian of the subsystem of the $^{73}$Ge magnetic nuclei in the germanium single crystal (the lattice sites are labeled by $k,j$) can be written as follows:

$$H = \sum_k H_k^Z + \sum_k H_k^Q + \sum_{j>k} H_{jk}^{dd} + H_1 + H_2. \tag{1}$$

Here, the first term stands for the interaction with the external magnetic field $H$. In the system of coordinates with the $z$-axis directed along the magnetic field, the nuclear Zeeman energy equals

$$H_k^Z = \gamma \hbar H I_{zk} \tag{2}$$

where $\gamma = 2\pi \times 1.493$ MHz/T is the $^{73}$Ge nuclear gyromagnetic ratio. We consider here the strong magnetic field approximation when quadrupole splittings and the energy of the magnetic dipole-dipole interaction are much less than the Zeeman splitting. In this case the second term in Eq.(1), corresponding to the energy of the nuclear quadrupole moment $Q = -0.173 \cdot 10^{-24}$ cm$^2$ [17] in the static lattice, equals

$$H_k^Q = \frac{e^2 Q V_{zz}(k)}{4I(2I-1)} \left[ 3I_{zk}^2 - I(I+1) \right], \quad (3)$$

where $V_{zz}(k)$ is the corresponding component of the EFG tensor (in units of proton charge $e$). The third term in Eq.(1) describes the secular part of the magnetic dipole-dipole interactions ($\mathbf{r}_{jk}$ is the vector connecting magnetic nuclei):

$$H_{jk}^{dd} = \gamma^2 \hbar^2 \frac{1}{r_{jk}^3} \left( 1 - \frac{3z_{jk}^2}{r_{jk}^2} \right) \left( 3I_{zj} I_{zk} - \mathbf{I}_j \mathbf{I}_k \right). \quad (4)$$

The nuclear spin-phonon interactions, linear and quadratic in displacements of atoms from their equilibrium positions, are represented by the last two terms in (1), respectively.

The EFG is a functional of the charge density in a crystal. As mentioned in the introduction, the static part of the EFG at lattice sites vanishes in the perfect germanium crystal. The random distribution of different germanium isotopes lowers the local lattice symmetry and induces fluctuations of the EFG. The corresponding fluctuations of the quadrupole energies and the magnetic dipole-dipole interactions contribute to the NMR linewidth and determine the lineshape of the NMR signal. An analysis of the lineshape may be carried out by using the statistical method[18][19][20] or the method of moments[21]. As we have shown earlier[14], in the germanium crystal with the natural isotope abundance (7.7 % of magnetic nuclei[22]), contributions to the linewidths of the NMR signals from the magnetic dipole-dipole interaction and the isotopic disorder have comparable values.

In the present work, we have measured NMR spectra in two magnetically diluted isotopically tailored germanium crystals containing ~0.1% of $^{73}$Ge isotope, in order to single out the quadrupole line broadening due to the isotopic disorder. From estimates of the contributions of the magnetic dipole-dipole interaction to the second moment of the resonance line, one may expect magnetic linewidths of the order of 1-10 Hz, thus the observed lineshape would be determined mainly by the random EFG. The samples differ essentially in the degree of the isotopic disorder (see Table I) that can be characterized by moments of the mass fluctuations $g_n = \frac{\langle \Delta m^n \rangle}{m^n}$, where $m = \sum_i c_i m_i$ is the average atomic mass, $c_i$ is the concentration of isotope $i$, and $\langle \Delta m^n \rangle = \sum_i c_i (m_i - m)^n$. It should be noted that in the earlier studies of nuclear acoustic resonance in single-crystal germanium with the natural $^{73}$Ge abundance[23], a Gaussian line shape for the $\Delta I_z = \pm 2$ transitions was detected in a magnetic field along the [110] direction at room temperature, having a second moment of 0.034 G$^2$ (the corresponding full width at half maximum (FWHM) equals 64 Hz and agrees with our data[14]). This width was explained in Ref.[23] as the sum of dipole-dipole and pseudo-exchange contributions. Effects of the isotopic disorder were neglected, and the treatment used for the pseudo-exchange is questionable (see below).

In the bond charge model, a primitive cell of the germanium crystal lattice (with the translation vectors $\boldsymbol{a}_1 = a[0\ 1/2\ 1/2]$, $\boldsymbol{a}_2 = a[1/2\ 0\ 1/2]$, $\boldsymbol{a}_3 = a[1/2\ 1/2\ 0]$, $a = 0.56575$ nm being the lattice constant) contains two germanium atoms (at the lattice sites $r_1 = (0\ 0\ 0)$, $r_2 = a(1\ 1\ 1)/4$) and four bond charges[15]. The EFG at the $^{73}$Ge nucleus located at the lattice site $L = 0$, $\lambda = 1$ ($L$ is the unit cell label, and $\lambda$ specifies the sublattice), can be written as follows (here and below Cartesian coordinates of vectors are labeled by Greek indices, the summation over the same Greek indices is always supposed):

$$V_{\alpha\beta} = \sum_{L,\lambda} \frac{Z_{L\lambda}(1-\gamma_{L\lambda})}{R^3(L0,\lambda 1)} \left( \frac{3 X_\alpha(L0,\lambda 1) X_\beta(L0,\lambda 1)}{R^2(L0,\lambda 1)} - \delta_{\alpha\beta} \right), \quad (5)$$

where

$$X_\alpha(LL',\lambda\lambda') = X_\alpha(L,\lambda) - X_\alpha(L',\lambda') \quad (6)$$

are relative coordinates of the lattice sites. In Eq.(5) $Z_{L\lambda}$ is either the atomic charge $Z_a$ ($\lambda = 1,2$) or the bond charge $Z_b$ ($\lambda = 3,4,5,6$), and $\gamma_{L\lambda}$ is the corresponding antishielding constant that depends on the distance $R(L0, \lambda 1)$. For the perfect lattice, it follows from the electroneutrality of a unit cell that $Z_a = -2Z_b = -2Z_0$. The value of the bond charge is determined by the overlapping of electron wave functions localized on the nearest neighbor Ge atoms. We assume, following Ref.[16] where dispersion of the Gruneisen parameters of different lattice modes in homopolar and heteropolar semiconductors was analyzed, that in the deformed lattice (under static or dynamic perturbations) the atomic charge equals one-half of the sum over four nearest neighbor bond charges with opposite signs, and the bond charge depends on the length of the corresponding bond $r_b$ as follows

$$Z_b(r_b) = Z_0 \left( \frac{r_b}{r_b^0} \right)^\rho, \quad (7)$$

where $r_b^0 = \sqrt{3}a/4$ is the length of the undeformed bond. We do not consider explicitly mechanisms of screening of the crystal electric field. The static dielectric constant of Ge, $\varepsilon = 16$, is determined mainly by electronic polarization[24]. We used the effective value of the bond charge $Z_0 = -0.4$ determined from the analysis of the phonon spectrum in Ref.[15]. This value is also not far from the average bond charge of about -0.58 calculated within the local-density approximation in the density-functional theory[25]. The exponent $\rho$ in Eq. (7) and the antishielding constants $\gamma_a = \gamma_\infty$ for all atomic charges and bond charges except the nearest neighbor bond charges, and $\gamma_b$ for the nearest bond charges, are treated as fitting parameters of the model.

The isolated mass defect at the lattice site $(L\lambda)$ with the mass difference $\delta m$ relative to the average atomic mass $m$ induces a totally symmetric local lattice deformation. In the case of a large distance $r(L0, \lambda 1)$ between the mass defect and the $^{73}$Ge nucleus at the site $L = 0$, $\lambda = 1$ under consideration, the effect of this local lattice deformation on the EFG at the magnetic nucleus may be considered as twofold: The redistribution of the charge density in the vicinity of the mass defect that contributes directly to the EFG defined in (5), and the strains in the vicinity of the magnetic nucleus which induce the EFG through the gradient-elastic tensor $S$[11]:

$$\delta V_{\alpha\beta}^{strain}(L\lambda) = S_{\alpha\beta\gamma\delta} e_{\gamma\delta}(L\lambda). \tag{8}$$

Here the labels $L\lambda$ ($\lambda = 1,2$) mean that the corresponding quantity is related to the mass defect at the site $(L\lambda)$. In the Ge crystal lattice composed of two cubic face centered Bravais sublattices, only the $S_{11}$, $S_{12}$ and $S_{44}$ components of the tensor $S$ (we use the Voigt notation in the crystallographic system of coordinates) are nonzero, and only two independent linear combinations ($S_{11} - S_{12}$ and $S_{44}$) can be obtained from experimental data on nuclear acoustic resonance (NAR)[23]. A volume change can not induce EFG in the cubic lattice. Thus, $S_{11} = -2S_{12}$, and we obtain

$$\delta V_{zz}^{strain}(L\lambda) = \frac{1}{2} S_{11}(2e_{zz}(L\lambda) - e_{xx}(L\lambda) - e_{yy}(L\lambda)), \qquad (9)$$

$$\delta V_{xy}^{strain}(L\lambda) = 2S_{44} e_{xy}(L\lambda). \qquad (10)$$

These relations are valid only for a homogeneous strain, such as found at large distances $R(L0, \lambda 1)$ from the lattice point defect.

Displacements of atoms from the perfect lattice sites at large distances $r$ from the mass defect $\delta m_i = m_i - m$ in the germanium lattice can be represented by[26]

$$u_x^i(r) = \frac{c_{11} + 2c_{12}}{32\pi c_{11}} a^2 \left(\frac{\delta a}{\delta m}\right) \delta m_i \frac{x}{r^3}[1 + \frac{c_{11} + c_{12}}{8c_{11}} \xi(1 - 9p_x^2 + 15p_x^4 - 15p_y^2 p_z^2) + \frac{c_{12}}{c_{11}} \xi(1 - 3p_x^2)], \qquad (11)$$

where $p_\alpha = x_\alpha / r$, $c_{ij}$ are the elastic constants ($c_{11}$=12.88, $c_{12}$ = 4.83, $c_{44}$ = 6.71 in units of $10^4$ J/cm$^3$)[15], $\xi = (c_{11} - c_{12} - 2c_{44})/c_{44}$ is the elastic anisotropy parameter (it is supposed that $|\xi| < 1$ in Eq. (11)), and $\delta a / \delta m$ is the lattice constant change per unit isotope mass. Within the elastic continuum approximation, the commonly used expression for displacements $u(r)$ may be obtained from (11) by setting $\xi = 0$, and $\frac{c_{11} + 2c_{12}}{c_{11}} = \frac{1+\sigma}{1-\sigma}$ where $\sigma$ is the Poisson's ratio. The components of the strain tensor to be used in Eqs.(9-10)

$$e_{\alpha\beta}^i(L\lambda) = \frac{1}{2}\left(\frac{\partial u_\alpha^i(R(L0,\lambda 1))}{\partial X_\beta(L0,\lambda 1)} + \frac{\partial u_\beta^i(R(L0,\lambda 1))}{\partial X_\beta(L0,\lambda 1)}\right) \qquad (12)$$

are easily obtained from (11).

The components of the displacement vector $\delta \mathbf{r}_p$ of the $p$-th neighbor of the isolated mass defect in the lattice site ($L\lambda$), having a charge $eq_p$, are denoted below as $\delta x_p$, $\delta y_p$, $\delta z_p$. The local lattice deformation induces a change of the electric field at the site $L = 0$, $\lambda = 1$ where we place a $^{73}$Ge nucleus. The change of the crystal electric field potential at the large distance $r$ from the mass defect, linear in $\delta \mathbf{r}_p$, can be written in the following invariant form

$$\delta V(\mathbf{r}) = -\sum_p eq_p [\frac{15}{2r^4}(\mathbf{r}\delta\mathbf{r}_p/r)(\mathbf{r}\mathbf{r}_p/r)^2 + O((r_p/r)^3)]. \qquad (13)$$

Let us denote the radial displacements of Ge atoms in the first coordination shell of the isolated impurity isotope with the unit mass defect $\delta m_i = 1$ as $\sqrt{3}\delta_a$, the radial displacements of the nearest bond charges as $\sqrt{3}\delta_b$,, the displacement vector of the bond charge with coordinates $(a/8)(1,3,3)$ as $(\delta_{b1}\ \delta_{b2}\ \delta_{b2})$. The displacements of other bond charges in this coordination shell can be obtained by application of symmetry operations. It follows from the results of calculations, described in Ref.[14], that displacements of more distant atoms and bond charges do not exceed 1/10 of the nearest neighbor displacements. The parameters of the local lattice deformation $\delta_a$, $\delta_b$, $\delta_{b1}$, $\delta_{b2}$ were calculated using the theory presented in Ref.[14], their values are given in Table II. The anharmonic force constants used in the calculations are discussed in the following section.

The anisotropic properties of a cubic lattice reveal themselves by different responses to the external magnetic field along either tetragonal or trigonal symmetry axes. In the first case ($H\|[001]$), we obtain from Eqs.(5) and (13) the EFG caused by the redistribution of the charge density in the vicinity of the mass defect in the following form

$$\delta V_{zz}^{ch,i}(L\lambda) = 90(-1)^\lambda (1-\gamma_\infty)\frac{a^2}{64}\delta m_i [Z_0\Delta + \frac{a}{8}\delta Z_b]V_5^{-2}(\boldsymbol{R}(L0,\lambda 1))/R(L0,\lambda 1)^6, \quad (14)$$

where as in the second case ($H\|[111]$):

$$\delta V_{z'z'}^{ch,i}(L\lambda) = 10\sqrt{3}(-1)^\lambda (1-\gamma_\infty)\frac{a^2}{64}\delta m_i [Z_0\Delta + \frac{a}{8}\delta Z_b][5V_5^0(\boldsymbol{R'}(L0,\lambda 1)) + \\ \sqrt{2}V_5^{-3}(\boldsymbol{R'}(L0,\lambda 1))]/R(L0,\lambda 1)^6. \quad (15)$$

Here $\Delta = 8\delta_a - \delta_b - 9\delta_{b1} - 6\delta_{b2}$, $V_5^{-2}(\boldsymbol{r}) = 42xyz(3z^2 - r^2)/r^5$,

$V_5^0(\boldsymbol{r}) = z(63z^4 - 70z^2r^2 + 15r^4)/r^5$, $V_5^{-3}(\boldsymbol{r}) = 7y(3x^2 - y^2)(9z^2 - r^2)/r^5$,

and the components of vectors $\boldsymbol{R'}$ are given with respect to the coordinates system $x'=(x-y)/\sqrt{2}$, $y'=(x+y-2z)/\sqrt{6}$, $z'=(x+y+z)/\sqrt{3}$ with $z'\|[111]$. We have included in Eqs. (14) and (15) additional contributions to the EFG induced by the changes $\delta Z_b = (4\delta_a/a)Z_0\rho$ of bond charges localized on the bonds connecting the mass defect with its nearest neighbors.

The Hamiltonian of the quadrupole interaction of the $^{73}$Ge nucleus at the site $L = 0$, $\lambda = 1$ in the isotopically disordered lattice can be represented as

$$H_k^Q = \frac{e^2Q}{4I(2I-1)}[3I_{zk}^2 - I(I+1)]\sum_{L\lambda,i}\delta V^i(L\lambda)c^i(L\lambda), \quad (16)$$

where $c^i(L\lambda)$ are random occupation numbers equal to unity if the site $(L\lambda)$ contains an atom with the mass $m_i$. Additive contributions to the fluctuations of the EFG from the distant mass defects are determined by local strains and charge redistributions. For $H\|[001]$ and $H\|[111]$, from Eqs.(9,11,14,15), we obtain, respectively,

$$\delta V^i_{zz}(L\lambda) = \delta V^{ch,i}_{zz} + \tfrac{1}{2} S_{11}[2e^i_{zz}(L\lambda) - e^i_{xx}(L\lambda) - e^i_{yy}(L\lambda)], \tag{17}$$

$$\delta V^i_{zz}(L\lambda) = \delta V^{ch,i}_{z'z'} + \tfrac{4}{3} S_{44}[e^i_{xz}(L\lambda) + e^i_{xy}(L\lambda) + e^i_{yz}(L\lambda)]. \tag{18}$$

The contributions to the EFG from the mass defects occupying sites in the first and second coordination shells of the magnetic nucleus were obtained as explicit functions of parameters of the local lattice deformation directly from the definition of the EFG in Eq.(5). In particular, mass defects in the first coordination shell of the magnetic nucleus do not affect the nuclear quadrupole moment in the magnetic field $H\|[001]$. In this case, the main contributions to the EFG are induced by mass defects in the second coordination shell. Neglecting shifts of the bond charges (see Table II) and correlations between deformations due to adjacent mass defects, we obtained for four equivalent defect positions of the type A ($[a/2\ a/2\ 0]$) and for eight equivalent positions of the type B ($[a/2\ 0\ a/2]$)

$$\delta V^i_{zz}(A) = -2\delta V^i_{zz}(B) = (-285 + 0.66\rho)\delta_a Z_0 (1-\gamma_\infty)\delta m_i / a^4. \tag{19}$$

The largest quadrupole splitting is induced in a magnetic field $H\|[111]$ by the mass defect at the nearest neighbor site $A_1$ ($[a/4\ a/4\ a/4]$) along the field direction (the contributions from the three other equivalent sites $B_1$ in the first coordination shell are three times smaller):

$$\delta V^i_{z'z'}(A_1) = -3\delta V^i_{z'z'}(B_1) = [(1100 - 215\rho)(1-\gamma_\infty) - (10510 - 788\rho)(1-\gamma_b)]\delta_a Z_0 \delta m_i / a^4. \tag{20}$$

In this case the magnetic nucleus is displaced from the tetrahedron center, and the EFG includes the contribution from the corresponding term in the crystal field potential of the perfect lattice. The diagonal EFG components along the [111] crystallographic axis induced by mass defects in the second coordination shell of the magnetic nucleus equal

$$\delta V^i_{z'z'}(A_3) = (-165 + 1.42\rho)(1 - \gamma_\infty)\delta_a Z_0 \delta m_i / a^4,$$

$$\delta V^i_{z'z'}(B_3) = (285 + 41.3\rho)(1 - \gamma_\infty)\delta_a Z_0 \delta m_i / a^4,$$

$$\delta V^i_{z'z'}(C_3) = (-60 - 21.4\rho)(1 - \gamma_\infty)\delta_a Z_0 \delta m_i / a^4,$$

where $A_3$, $B_3$, $C_3$ correspond to three equivalent sites of the type [$a/2$ $a/2$ 0], three sites of the type [$-a/2$ $-a/2$ 0], and six sites of the type [$a/2$ $-a/2$ 0], respectively.

The simplest signature of the EFG distribution is its mean square value

$$<\delta V^2> = \sum_i c_i \sum_{L\lambda}[\delta V^i(L\lambda)]^2 = m^2 g_2 \sum_{L\lambda}\left[\frac{\delta V^i(L\lambda)}{\delta m_i}\right]^2 \qquad (21)$$

that determines frequency shifts of nuclear transitions. In particular, the mean square frequency shift of the $\pm 3/2 \leftrightarrow \pm 1/2$ nuclear transition equals

$$(\Delta \nu_Q)^2 = <(\nu_{\pm 3/2 \leftrightarrow \pm 1/2} - \nu_0)^2> = \left[\frac{3e^2 Q}{4\pi \hbar I(2I-1)}\right]^2 <\delta V^2>, \qquad (22)$$

where $\nu_0 = \gamma H / 2\pi$ is the NMR frequency.

Due to a large number of random configurations of mass defects that induce quadrupole shifts of nuclear sublevels of comparable values in a strong magnetic field, the isotopic structure of the NMR line is smoothed out, and the NMR absorption lineshape may be approximated by a function with one maximum at the resonance frequency $\nu_0$. Some information about this function may be obtained from estimates of its second ($M_2$) and fourth ($M_4$) moments. Contributions to these moments due to the random EFG are proportional to $<\delta V^2>$ (see Eq. (21)) and $<\delta V^4>$, respectively, where

$$<\delta V^4> = <(\sum_i c^i(L\lambda)\delta V^i(L\lambda))^4> = 3(<\delta V^2>)^2 + (g_4 - 3g_2^2)m^4 \sum_{L\lambda}\left[\frac{\delta V^i(L\lambda)}{\delta m_i}\right]^4. \quad (23)$$

It is well known that the fourth and the second moments of the resonance signal are connected by the simple relation $M_4 = 3(M_2)^2$ for a Gaussian. The calculated ratios

$$\kappa_Q = \frac{M_4}{3M_2^2} = \frac{<V^4>}{3(<V^2>)^2}$$ (see Table I) do not differ much from unity in the samples

studied in this work. The sum of Gaussians corresponding to magnetic dipole transitions $I_z \leftrightarrow I_z \pm 1$ with widths determined by the mean square $<\delta V^2>$ of the EFG fluctuations at the $^{73}$Ge nuclei is used below to describe the isotopically-induced quadrupole broadening of resonance lines:

$$f(\nu) = \frac{|<\tfrac{1}{2}|I_x|-\tfrac{1}{2}>|^2 \delta(\nu - \nu_0) + \sum_{m=3/2}^{m=9/2} \frac{2|<m|I_x|m-1>|^2}{[\pi(2m-1)^2 \Delta\nu_Q^2/2]^{1/2}} \exp\left(-\frac{(\nu - \nu_0)^2}{[(2m-1)^2 \Delta\nu_Q^2/2]}\right)}{\sum_{m=-7/2}^{m=9/2} |<m|I_x|m-1>|^2}$$

(24)

Here the transition $+1/2 \leftrightarrow -1/2$ is represented by a $\delta$-function because it is not affected by the EFG. To compare it with the experimental results, the calculated lineshape of the function determined in Eq.(24) must be convoluted with the form-function describing additional broadening of the NMR line due to magnetic interactions linear in nuclear spin operators such as a residual inhomogeneity of the static magnetic field within the sample.

## III. PARAMETERS OF THE MODEL

The quadrupole broadening of the NMR line effected by the isotopic disorder is determined by parameters that can be divided into two sets. The first set includes parameters of anharmonic interactions in the crystal lattice. These parameters are used in the calculations of the local lattice deformations induced by the isolated impurity isotope in the homogeneous crystal. The second set includes antishielding constants for the nearest bond charges ($\gamma_b$) and all other distant point charges ($\gamma_\infty$), introduced in our model of the quadrupole interaction, and the exponent $\rho$ that determines the changes of bond charges with the bond length. Both sets of parameters may be obtained in the framework of the bond charge model[15][16] from the theoretical analysis of independent additional experimental data.

The parameters of anharmonic interactions can be fitted to the measured temperature dependence of the expansion coefficient of germanium[27], and the isotopic effects on the lattice constant and its temperature dependence[28][29] investigated recently with the use of the X-ray standing wave[30][31] and backscattering[32] techniques. The lattice constant at different temperatures and its dependence on the isotopic composition can be calculated from the minimum condition of the crystal lattice free energy[9][33]. The full symmetric deformation of the germanium crystal lattice is determined by the trace of the strain tensor

$$e_{\alpha\alpha} = -(c_{11} + 2c_{12})^{-1} T_{\alpha\alpha} / v_0, \tag{25}$$

where $v_0$ is the volume of the unit cell, and $T_{\alpha\beta}$ is the internal stress tensor, a functional of spectral densities of displacement-displacement correlation functions[9] which depends on the temperature and the concentrations of the different Ge isotopes. In the framework of the anharmonic bond charge model[16], the deformation tensor can be represented by a linear function of the charge redistribution parameter $\rho$ defined in Eq. (7) and of the parameters that characterize the linear scaling of the central non-Coulombic atom-atom ($m_{a1-a2}$,) and atom-bond charge ($m_{ab}$) interactions, and of the Keating bond-bending interaction ($m_b$) with the relative change in the distance between the nearest neighbor Ge atoms, Ge atom and the bond charge, and between the bond charges, respectively. We have been able to describe the temperature dependence of the germanium thermal expansion (see Fig. 1a) and of the lattice constant changes in the isotopically enriched germanium crystals (Fig. 1b) using the spectral densities of the correlation functions for relative dynamic displacements of Ge atoms and bond charges, obtained in the framework of the bond charge model in our previous work[34], and the following values of the parameters defined above: $m_{a1-a2}$ = - 12 (- 12.73), $m_{ab}$= 8 (12.27), $m_b$= - 5 (- 1.18), $\rho$ = - 12 (12.55) (values of the corresponding parameters from Ref.[16] are in the parentheses). The calculated temperature dependence of the difference between linear coefficients of thermal expansion in the $^{70}$Ge crystal and in the natural Ge crystal divided by the difference between average atom masses (this difference equals 2.59 atomic units), displayed by the dotted curve in Fig.1b, agrees satisfactorily with the experimental data of Ref.[29] in the range of 30-100 K. However, the experimental data above 100 K in Fig.2 of Ref.[29] are not correct. The parameters $\rho$ of the charge redistribution in the present work and in Ref.[16] have been defined with different signs. In our model, the bond charge increases with decreasing bond length, such a behaviour correlates with the overlapping of atomic wave functions and allows us to determine reasonable values of antishielding factors (see below). Using the expression $|Z_0| = Z/\varepsilon$ for the bond charge[15][16] we also find an increase of $|Z_0|$ with decreasing bond length since the latter induces a decrease in $\varepsilon$[35].

The parameters of the interaction between the $^{73}$Ge quadrupole moment and the crystal electric field can be obtained from experimental data on nuclear spin-lattice relaxation and NAR. In our previous work[34], we were able to interpret the temperature dependence of the spin-lattice relaxation time in Ge single crystals in the framework of the anharmonic bond charge model by introducing two different antishielding constants: $\gamma_b$ for bond charges, which are the nearest neighbors of the $^{73}$Ge nucleus, and $\gamma_\infty$ for all other charges localized on atoms and bonds. Additional information about the shielding constants may be obtained from the measured components of the gradient-elastic tensor[23], the corresponding analysis of this tensor in the framework of the bond charge model is given below.

The displacements of atoms and bond charges $u_\alpha(L\lambda)$ under uniform crystal lattice deformation are determined by elastic strain tensor components $e_{\alpha\beta}$ and sublattice internal displacements $w(\lambda)$:

$$u_\alpha(L\lambda) = e_{\alpha\beta} X_\beta(L\lambda) + w_\alpha(\lambda), \qquad (26)$$

where internal displacements are linear in components of the strain tensor:

$$w_\alpha(\lambda) = \Gamma_{\alpha\beta\gamma}(\lambda) e_{\beta\gamma}. \qquad (27)$$

The internal displacement $w(\lambda)$ equals zero at the sites of the sublattice $\lambda$ which are centers of inversion. The bond charges in the perfect Ge crystal are at centers of inversion, and their displacements under a uniform stress are determined entirely by components of the macroscopic strain tensor only. At the Ge atom sites with tetrahedral symmetry, the components $\Gamma_{xyz}(1) = -\Gamma_{xyz}(2) = a\zeta/8$ of the tensors $\Gamma(1)$, $\Gamma(2)$ in the crystallographic system of coordinates are different from zero ($\zeta$ is the single dimensionless parameter of the internal displacements of atomic sublattices). The values of these components were found from studies of the lattice dynamics using the longwave method. The parameters of the quasi-uniform lattice deformation induced by the longwave acoustic vibration with the wave vector $q \to 0$ and the polarization vector $e(q\eta/\lambda)$ can be written as follows ($\eta$ is the label of acoustic branches of the phonon spectrum):

$$e_{\alpha\beta} = \frac{1}{2}\left[q_\alpha \mathrm{Re}\, e_\beta(0\eta/\lambda) + q_\beta \mathrm{Re}\, e_\alpha(0\eta/\lambda)\right] m(\lambda)^{-1/2},$$
$$w_\alpha(\lambda) = m(\lambda)^{-1/2} q_\beta \frac{\partial}{\partial q_\beta} \mathrm{Im}\, e_\alpha(q\eta/\lambda)\big|_{q=0}. \qquad (28)$$

From calculations of imaginary parts of polarization vectors of acoustic modes at different points near the Brillouin zone center, we obtained the value of the dimensionless parameter $\zeta = 0.523$. A listing of the various values of the strain parameter $\zeta$ found in the literature is given in Ref.[36]. For our calculations we took the typical value $\zeta = 0.52$.

According to Eqs.(8-10), we can express the nonzero components of the gradient-elastic tensor in terms of antishielding factors for the nearest and distant charges, parameters of internal displacements and lattice sums:

$$S_{11} = \sum_{L,\lambda} V_{zz,z}(L\lambda)Z(L0,\lambda 1),$$

$$S_{44} = \sum_{L,\lambda} \left\{ \frac{1}{2} \left( V_{xy,x}(L\lambda)Y(L0,\lambda 1) + V_{xy,y}(L\lambda)X(L0,\lambda 1) \right) + V_{xy,z}(L\lambda)\left( \Gamma_{zxy}(\lambda) - \Gamma_{zxy}(1) \right) \right\}. \quad (29)$$

With the third order force constants $V_{\alpha\beta,\gamma}(L\lambda) = \partial V_{\alpha\beta}/\partial X_\gamma(L0,\lambda 1)$, calculated in the framework of the anharmonic bond charge model, we obtain

$$S_{11} = -1.161(1-\gamma_b) + 0.203(1-\gamma_\infty), \quad (30)$$

$$S_{44} = -\tfrac{1}{2}(1-\tfrac{5}{2}\zeta)S_{11} - 0.29(1-\gamma_b)(1-\zeta)\rho + 0.3\zeta(1-\gamma_\infty). \quad (31)$$

The absolute values $|S_{11}| = 7.08$ and $|S_{44}| = 26.04$ in units of $10^{24}$ см$^{-3}$ of the gradient-elastic tensor components in Ge crystals were obtained from the NAR experiments in Ref.[23], the spin-lattice relaxation times at different temperatures were measured in Refs.[34,37,38]. By fitting to the experimental data the simulated rates of the quadrupole NMR-relaxation (the corresponding theory was derived in Ref.[34]) and the components of the gradient-elastic tensor, we obtained $1-\gamma_b = 9$, $1-\gamma_\infty = 86.2$ (for the fixed value of $\rho = -12$). The values of the gradient-elastic tensor components, as calculated with Eqs.(30) and (31), are $S_{11} = +7.08$, $S_{44} = +17.8$ (in units of $10^{24}$ cm$^{-3}$). It was shown in Ref.[39] that simple ionic models cannot explain the observed ratio $|S_{11}|/|S_{44}|$ in the tetrahedral III-V semiconductors. The discrepancy between experimental and calculated values is due to the covalent character of interatomic bonds. In the framework of the anharmonic bond charge model, this effect is partly accounted for by the charge redistribution (7) that accompanies shifts of Ge atoms from their equilibrium positions. The simulated spin-lattice relaxation times $T_1$ agree fairly well with the measured data (see Fig. 2).

The parameters of the anharmonic and quadrupole interactions presented above were used to calculate the second (Eq. (21)) and fourth (Eq. (23)) moments of the EFG distribution induced by the isotopic disorder in the samples studied in this work. The corresponding parameters of the quadrupole broadening of NMR lines are compared with the experimental data in Table I.

It should be noted that it is possible to obtain almost the same absolute values of $S_{11}$ and $S_{44}$ components of the gradient-elastic tensor (with opposite signs) and a satisfactory description of relaxation rates using a positive value of the parameter $\rho = 12.55$ (from Ref.[16]) and the same value of $1-\gamma_b$, but with $1-\gamma_\infty = 16$ that might be compared with the value 7.8 of this factor used in Ref.[23] together with the Ge charge +4. However, a comparison of phenomenological parameters of models based on different physical grounds cannot be justified.

## IV. Experimental results and discussion

NMR spectra of $^{73}$Ge have been measured in two single crystals (6x6x6 mm$^3$) of highly pure germanium (carrier concentration $n_{carr} \sim 10^{12}$ cm$^{-3}$ at 290 K) with different isotopic compositions (see Table I). The density of dislocations in the Ge crystals was less than $10^3$ cm$^{-2}$ as followed from the decoration-etching test[40].

The measurements were performed at room temperature ($T = 300$ K) with a Bruker ASX 500 NMR spectrometer that operated at the resonance frequency of the $^{73}$Ge isotope ($\nu_{Ge} = 17.44$ MHz in a field $H = 11.7$ T). The free induction decay signal (FID) of the $^{73}$Ge isotope was detected after applying a single exciting pulse with a duration of ~ 5 μsec, corresponding to a $\pi/3$ exciting pulse. The spectral width of the exciting pulse was enough to excite uniformly the whole spectrum of $^{73}$Ge. The experimental procedure was repeated many times after the time interval t ~ $T_1$ (300 K) = 11 sec[34] until the required signal to noise ratio (>50) had been achieved. Approximately 4000 accumulations were necessary for getting the required performance. The spectra were obtained by a Fourier-transformation of the accumulated FID. The experimental results are shown in Figs. 3a,b and 3c,d for *H*||[001] and *H*||[111], respectively. Before carrying out the measurements, the inhomogeneity of the static magnetic field was minimized, with the standard shimming procedure, to 0.2 ppm within the sample, using the NMR signal of $^{39}$K nuclei ($\nu_K$ = 23.33 MHz) in a water solution of KBr placed in a capsule of the same shape as the crystals to be studied. The resulting NMR line of $^{39}$K, that could be well represented by a Lorentzian with the FWHM $\Delta\nu_{app}$ = 7 Hz, scaled in width like the ratio of ($\nu_{Ge}/\nu_K$), was used to estimate the parameters of the form-function, which determined the spectrometer resolution.

In Figs. 3a - 3d, the NMR lines in the $^{70}$Ge, $^{70/76}$Ge samples are compared with the simulated signals, corresponding to the function of Eq.(24) convoluted with a Lorentzian profile whose width $\Delta\nu_{app}$ = 5.2 Hz was determined from the spectrometer resolution. The observed lineshapes depend strongly on the magnetic field orientation and are characterized by a plateau whose width increases with increasing degree of isotopic disorder and with the rotation of the magnetic field in the (0 1 1) plane from the tetragonal to the trigonal symmetry axis.

The parameters of the quadrupole broadening $\Delta\nu_Q$ of the NMR lines extracted from the fits to the measured spectra for orientations of the external magnetic field along the [001] and the [111] crystallographic directions are compared in Table I. The calculated widths agree very well with the widths of the fits of the simulated signals to the experimental spectra for $H\|[001]$. For $H\|[111]$, the relative differences between experimental and theoretical values of $\Delta\nu_Q$ are similar to the relative difference between the measured and calculated values of the $S_{44}$ component of the gradient-elastic tensor (see preceding section), and must have their origin in the crude treatment of the covalent bond contributions to the EFG in the framework of the bond charge model. As seen in Fig. 4, the measured linewidths $\Delta\nu_Q$ are proportional to the square root of the second moment of the isotopic mass distribution. This fact justifies the use of Gaussian lineshapes to describe individual transitions between the nuclear Zeeman sublevels. The ratio of the linewidths $R = \dfrac{\Delta\nu_Q(H\|[111])}{\Delta\nu_Q(H\|[001])}$, which determines the anisotropy of the lineshape due to the isotopic disorder, should not depend on the sample. In the $^{70}$Ge and $^{70/76}$Ge samples, the mesured values of $R$ equal 3.60 and 3.12, respectively (preliminary data on NMR spectra in the $^{70}$Ge sample, presented in Ref.[14], yielded a lower $R$ value because of misorientation of the sample). The average $R = 3.36$ coincides with the $R$ value extracted from the NMR spectra of a natural Ge crystal ($g_2 = 5.88\times10^{-4}$)[14], where the concentration of magnetic nuclei (c($^{73}$Ge) = 7.73 %) is high and the magnetic dipolar broadening becomes important. In this case, we used the same superposition of Gaussians [Eq.(24)] to describe the measured NMR signals, but the second moment of each transition (including the 1/2 ↔ -1/2 transition) was taken to be the sum of the magnetic broadening term $<\Delta\nu^2>_M$ and the quadrupolar one:

$$<\Delta\nu^2>_{m\to m-1} = (2m-1)^2 \Delta\nu_Q^2/4 + <\Delta\nu^2>_M, \quad (m = 9/2, 7/2, \ldots, -7/2) \qquad (32)$$

Quadrupolar and magnetic contributions to the linewidths of comparable magnitudes were found by fitting the model lineshape to the observed ones. The corresponding values of $\Delta\nu_Q$ are presented in Fig. 4. In natural Ge crystals, the calculated second moments of the resonance lines corresponding to magnetic dipole-dipole interactions agree satisfactorily with the experimental values of $<\Delta\nu^2>_M$ [14]. Thus, the magnetic dipole-dipole interaction and the quadrupole interaction with the random EFG due to the intrinsic isotopic disorder are the most important sources of NMR line broadening in Ge crystals.

## V. SUMMARY

The isotopic disorder in crystals primarily affects the vibrations of the crystal lattice. Complementary information about random lattice strains induced by isotopic disorder in semiconductors may be obtained from studies of NMR in magnetic isotopes. NMR techniques may be also used in nondestructive investigations of isotopically engineered semiconductor structures[41,42]. The results of the present work demonstrate that NMR spectral lines of nuclei with quadrupole moments have a characteristic shape in crystals with isotopic disorder.

The isotopic disorder lifts the cubic symmetry of the perfect Ge lattice and produces random electric field gradients at the lattice sites. We applied a unified theoretical approach based on the adiabatic bond charge model[15] to describe both, lattice deformations induced by mass defects in the anharmonic crystal and random electric fields accompanying these deformations in the isotopically mixed Ge crystals. All model parameters used in calculations of the NMR lineshapes were obtained from the analysis of experimental data found in the literature for the thermal expansion of the crystal lattice, the gradient-elastic tensor, the temperature dependence of the nuclear spin-lattice relaxation, and the isotopic effect on the lattice constant of germanium. A reasonable agreement is obtained between experimental NMR spectra and simulated absorption lines for different orientations of the magnetic field. The quadrupole broadening of resonance lines due to the isotopic disorder reaches its maximum and minimum for the magnetic field along the trigonal and tetragonal symmetry axis of the Ge crystal lattice, respectively, contrary to the case of impurities in alkali halide crystals, where these extremal directions of the external magnetic field correspond to a minimum and maximum of the NMR linewidth[11,12]. In both cases, the largest effect is observed when the magnetic field is directed along the bonds connecting the magnetic nucleus with its nearest neighbors.

In conclusion, it should be noted that, in contrast to the specific quadrupolar broadening of the NMR transitions, the isotopic structures of narrow optical f-f transitions in impurity rare earth ions in crystals with the intrinsic isotopic disorder have been observed[9]. The crystal field energies of the localized f-electrons are determined mainly by their interaction with the nearest neighbor ions, and the isotopic structure is resolved due to the prevailing role of the isotopic disorder in the first coordination shell of the rare earth ion. In an external magnetic field, similarly to the broadening of the NMR lines discussed here, the isotopic structure may be smoothed out because different nearest neighbor sites become nonequivalent.

It has been recently reported that the edge emission of silicon is strongly broadened by the isotopic disorder[43]. The theoretical treatment of this disorder bears a resemblance to that reported for NMR lines in the present work.

## ACKNOWLEDGEMENTS

The authors would like to thank E.E. Haller for the isotopically tailored germanium samples. This work was partly supported by the CRDF grant 2274.

Table I. Sample characteristics and parameters of the NMR lineshapes measured at room temperature. Columns $\Delta\nu_Q$ contain fitted and calculated (in brackets) quadrupole widths. Columns $\kappa_Q$ display the calculated ratios of the fourth moment to three-times the square of the second moment of the distribution functions of the quadrupole shifts.

| Sample | | | | | $H \parallel [001]$ | | $H \parallel [111]$ | |
|---|---|---|---|---|---|---|---|---|
| $m$ (at. units) | Isotope content (%) | $10^4\, g_2$ | $10^8\, g_4$ | | $\Delta\nu_Q$ (Hz) | $\kappa_Q$ | $\Delta\nu_Q$ (Hz) | $\kappa_Q$ |
| $^{70}$Ge 70.03 | $^{70}$Ge – 96.3<br>$^{72}$Ge – 2.1<br>$^{73}$Ge – 0.1<br>$^{74}$Ge – 1.2<br>$^{76}$Ge – 0.3 | 0.775 | 27.9 | | 10.28 (9.5) | 4.13 | 37.04 (24.2) | 1.62 |
| $^{70/76}$Ge 73.20 | $^{70}$Ge – 43<br>$^{72}$Ge – 2<br>$^{73}$Ge – 0.1-0.2<br>$^{74}$Ge – 7<br>$^{76}$Ge – 48 | 15.32 | 263.5 | | 46.4 (44.2) | 0.87 | 144.7 (111.6) | 0.97 |

Table II. Parameters of the local lattice deformation induced by the impurity isotope center at the site (0 0 0) in the Ge crystal at 300 K (in units of $10^{-7}$ nm per unit mass difference)

| Charge coordinates | Displacement |
|---|---|
| bond charge  $a$[1/8 1/8 1/8] | [$\delta_b$ $\delta_b$ $\delta_b$]   $\delta_b = -0.218$ |
| atom   $a$[1/4 1/4 1/4] | [$\delta_a$ $\delta_a$ $\delta_a$]   $\delta_a = -1.064$ |
| bond charge  $a$[1/8 3/8 3/8] | [- 0.10 - 0.11 - 0.11] |
| atom   $a$[0 1/2 1/2] | [- 0.133 - 0.153 -0.153] |

**Figure captions**

**FIG. 1. The thermal expansion coefficient of Ge (a), and the temperature dependence of the lattice constant change per unit atomic mass (b). Experimental data are shown by triangles[27], circles[31] and squares[32], solid curves are obtained from calculations (see also Ref.[33]). The dotted curve in Fig.1b represents the derivative of the solid curve (the corresponding scale is given on the right y-axis).**

**FIG. 2. Dependence of the spin-lattice relaxation time on temperature in Ge single crystals with natural isotope abundance. The solid line represents the results of calculations; the black and open circles correspond to experimental data from Ref.[34] and Refs.[37,38], respectively.**

**FIG. 3. NMR spectra of $^{73}$Ge in single crystals $^{70}$Ge and $^{70/76}$Ge. The experimental data are represented by circles; the solid lines correspond to the calculations.**

**FIG. 4. Measured (filled symbols) and calculated (open symbols) NMR linewidths as functions of the isotopic disorder parameter $g_2$ in magnetic fields $H\|[111]$ (squares) and $H\|[001]$ (circles).**

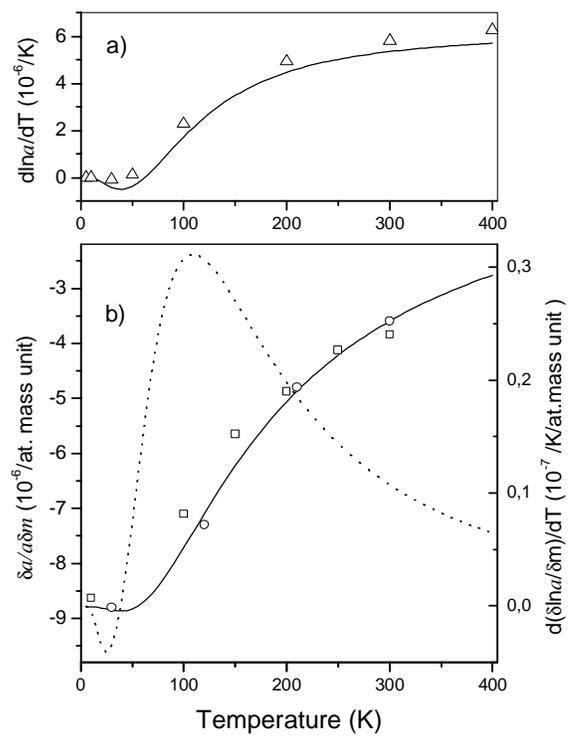

Fig.1

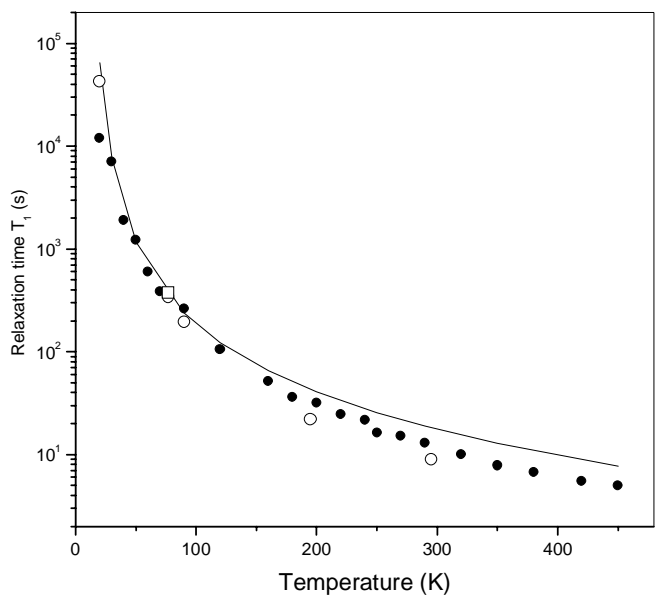

Fig.2

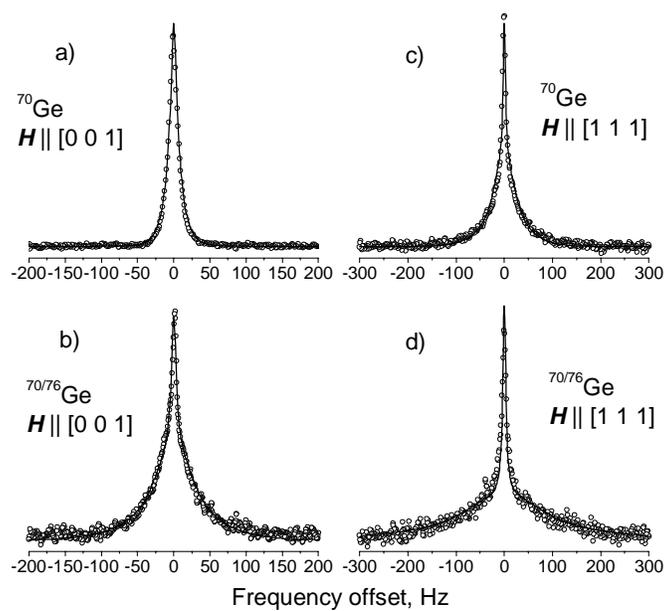

Frequency offset, Hz

Fig.3

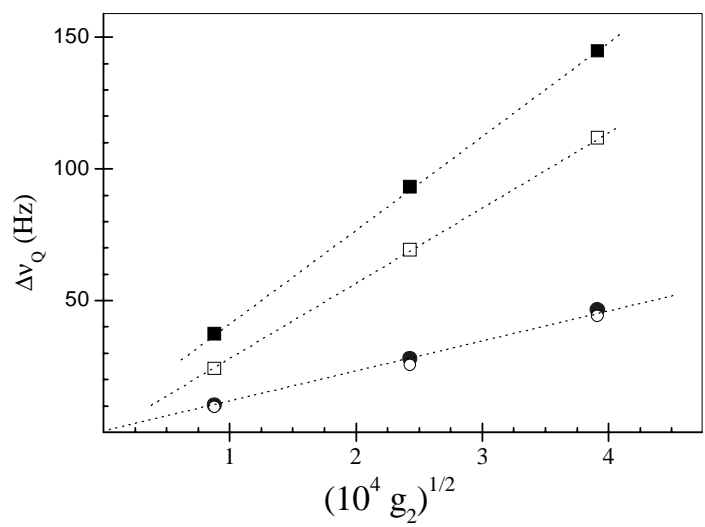

Fig.4